%% file: main.tex
\documentclass[sigconf]{acmart}
\usepackage{subfigure}

\AtBeginDocument{%
  \providecommand\BibTeX{{%
    \normalfont B\kern-0.5em{\scshape i\kern-0.25em b}\kern-0.8em\TeX}}}

\setcopyright{none}
\copyrightyear{2018}
\acmYear{2018}
\acmDOI{XXXXXXX.XXXXXXX}

\acmConference[Conference acronym 'XX]{Make sure to enter the correct
  conference title from your rights confirmation emai}{June 03--05,
  2018}{Woodstock, NY}
\acmPrice{15.00}
\acmISBN{978-1-4503-XXXX-X/18/06}

\begin{document}

\title{Large-Scale Assessment of Labour Market Dynamics in China during the COVID-19 Pandemic}


\author{Ying Sun$^{1}$, Hengshu Zhu$^{2}$, Hui Xiong$^{1}$}
\affiliation{
 \institution{$^1$The thrust of Artificial Intelligence, The Hong Kong University of Science and Technology, Guangzhou, 511455, China\\ $^2$Baidu Talent Intelligence Center, Baidu Inc., Beijing, 100085, China}
 \country{}
 }
\affiliation{yings@ust.hk, zhuhengshu@gmail.com, xionghui@ust.hk
 \country{}}

\renewcommand{\shortauthors}{Ying, et al.}

\begin{abstract}

The outbreak of the COVID-19 pandemic has had an unprecedented impact on China's labour market,  and has largely changed the structure of labour supply and demand in different regions. It becomes critical for policy makers to understand the emerging dynamics of the post-pandemic labour market and provide the right policies for supporting the sustainable development of regional economies. To this end, in this paper, we provide a data-driven approach to assess and understand the evolving dynamics in regions' labour markets with large-scale online job search queries and job postings. In particular, we model the spatial-temporal patterns of labour flow and labour demand which reflect the attractiveness of regional labour markets. Our analysis shows that regional labour markets suffered from dramatic changes and demonstrated unusual signs of recovery during the pandemic. Specifically, the intention of labour flow quickly recovered with a trend of migrating from large to small cities and from northern to southern regions, respectively. Meanwhile, due to the pandemic, the demand of blue-collar workers  has been substantially reduced compared to that of white-collar workers. In addition, the demand structure of blue-collar jobs also changed from manufacturing to service industries. Our findings reveal that the pandemic can cause varied impacts on regions with different structures of labour demand and control policies. This analysis provides timely information for both individuals and organizations in confronting the dynamic change in job markets during the extreme events, such as pandemics. Also, the governments can be better assisted for providing   the right policies on job markets in facilitating the sustainable development of regions' economies.

\end{abstract}



\maketitle
\input{data/intro}

\input{data/experiment}

\input{data/discussion}
\input{data/methods}

\bibliographystyle{ACM-Reference-Format}
\bibliography{main}

\end{document}

%% file: data/intro.tex
\section{Introduction}
In the past decades, the rapid growth of the Chinese economy has benefited tremendously from its prosperous labour market~\cite{fang2008impacts}. Indeed, the sustainable development of regional economics has been largely determined by the level of talent attraction in the region~\cite{hanushek2000schooling,hendricks2002important,bonaventura2021predicting}. However, the labour market is dynamic in nature and is vulnerable and sensitive to the policy change and the complicated social  environment~\cite{pries2005hiring,guirao2017high}. The outbreak of the COVID-19 pandemic has had an unprecedented impact on the labour markets worldwide~\cite{albanesi2021effects,brodeur2021literature,lee2021impact}. The structure of labour supply and demand in different regions of China has also been drastically changed by this pandemic. To this end, in this paper, we provide a data-driven fine-grained analysis to assess the regional labour market dynamics in China during the COVID-19 pandemic, which can benefit both individuals and organizations in confronting the uncertainty brought by the change of policy and society in the post-pandemic times. Meanwhile, this analysis can provide timely information to governments for better real-time policy-making in the macro regulation of labour market sustainable development.

Indeed, most of existing studies for labour market analysis are based on the census and survey data~\cite{boustan2010effect,schwandt2019unlucky,fortin2021labor}. However, due to the high cost of data collection, these data cannot support the fine-grained and timely analysis of regional labour market dynamics. For example, it is infeasible to collect sufficient regular data to capture the monthly or quarterly city-level labour migrants during the pandemic. Fortunately, with the rapid prevalence of online search engine and recruitment services, it is possible for us to study this problem with alternative large-scale data sources. Indeed, the search engine queries can naturally reflect users' demand and intention~\cite{ginsberg2009detecting,cousins2020regional}, which are widely used for crowd behavior profiling. The rationale behind is that when people have job transfer intention, many of them will seek for job information through the online search engines. Therefore, by mining the cross-city job seeking behaviors in the search engine queries, we can collect the labour flow intention data in a cost-effective way. Meanwhile, the massive job posting data in the online recruitment services can be naturally used for understanding the fine-grained labour demand structure and distribution in the labour markets of different regions.

The data used in this paper contain more than 2 billion online job search queries and 400 million job postings in China, across a time span of 21 months from September 2019 to June 2021 (see details in Methods). By using a combination of data mining approaches, we can effectively model the spatial-temporal patterns of labour flow and labour demand which reflect the attractiveness of regional labour market. Specifically, we first build a city-level labour flow intention graph by mining the spatial-temporal information in job search queries. Based on this graph, we can discover the overall labour agglomeration patterns in China, such as communities, black holes and volcanoes~\cite{hong2015detecting,li2010detecting}, by using link analysis and modularity-based clustering algorithms~\cite{newman2006modularity}. Then, based on the Hyperlink-Induced Topic Search (HITS) algorithm~\cite{kleinberg1998authoritative}, we can capture the temporal changes of regional labour market attractiveness from the labour flow intention during different periods of the pandemic, including lockdown (2020Q1), wide reopen (2020Q2), and recovery under regular pandemic control (after 2020Q2). Our analysis shows that the regional labour markets suffered from dramatic changes and demonstrated unusual signs of recovery during the pandemic. Specifically, the intention of labour flow quickly recovered with a trend of deviating from large to small cities and from northern to southern regions, respectively. Furthermore, with the text-clustering method~\cite{beil2002frequent}, we can classify job postings published in different regions into representative labour demands (e.g., blue collar and white collar) and analyze the change of regional labour demand during the pandemic. It is observed that the pandemic sharply decreased the labour demand of blue collar workers relative to that of white collar. In particular, the demand structure of blue collar jobs also changed from manufacturing to service industries. Our findings reveal that the pandemic might cause varied impacts on regions with different structures of labour demand and control policies.

%% file: data/experiment.tex

\vspace{-2mm}
\section{Results}
To capture the labour market dynamics during the pandemic, we propose to construct a labour flow intention graph, where nodes denotes cities and edges indicate labour flow intention between corresponding cities. Specifically, we exploited four classic link analysis metrics for measuring the labour market dynamics, including two local metrics \emph{Inflow} and \emph{Outflow}, and two global metrics \emph{Authority} and \emph{Hub} obtained by HITS algorithm. In particular, both \emph{Inflow} and \emph{Authority} indicate the strength of labour importation, while both \emph{Outflow} and \emph{Hub} indicate the strength of labour exportation.\footnote{Additional maps, terms, and figures can be found in supplementary information.}

\vspace{-1.5mm}
\subsection{The overview of labour flow intention}
\vspace{-0.5mm}
Based on the labour flow intention graph, we first sketch the characteristics of regional labour markets in China from 2019Q4 to 2021Q2. Generally, our analysis reveals that the labour flow intention is highly correlated with regional economic development.

First, \textbf{developed cities have more active labour flow intentions.} Specifically, we can observe that all the metric scores distribute unevenly among cities and are strongly correlated with regional GDP~\footnote{Data source: https://www.ceicdata.com/}(as shown in Table~\ref{tab:corr_gdp}).
This indicates that the developed cities usually have higher labour importation and exportation compared with other cities. Consequently, the distribution of labour flow intention is extremely imbalanced between eastern and western China. In western cities, few \emph{Inflow} nor \emph{Outflow} intentions exist, implying the inactive labour flow caused by their low development level and inconvenient transportation.

Second, \textbf{regional labour agglomeration exists in China.} Specifically, the modularity-based city clustering results (see Methods for details) indicate that cities have formed hierarchical communities of labour flow intention. Indeed, some clusters are provinces (e.g., Shandong province), reflecting inner-province labour exchange, while some clusters consist of multiple provinces (e.g., Jiangsu-Anhui-Zhejiang-Shanghai area), reflecting inter-province labour exchanges. Indeed, by adjusting the resolution, the detected communities can be further split or merged, reflecting the hierarchy of labour flow intention. 
In each community, most cities are \emph{Volcanoes} (i.e., the \emph{Outflow} score is larger than \emph{Inflow} score), while only a few cities are \emph{Black holes} (i.e., the \emph{Inflow} score is larger than \emph{Outflow} score), implying labours were hierarchically gathering in a few big cities. 

\begin{table}[t]
	\centering
    \caption{\textbf{The correlation between labour flow intention and regional GDP in 2020.} {\small For each metric, we calculated Pearson, Spearman, and Kendall correlation correlations.}}
	\label{tab:corr_gdp}
	\vspace{-2mm}
    \begin{tabular}{lllllll}
		\toprule
		Score & Pear. & $p$-value & Spear. & $p$-value & Kend. & $p$-value \\
	    \midrule
	    Inflow & 0.950 & $<0.001$ & 0.896 & $<0.001$ & 0.737 & $<0.001$ \\
	    Outflow & 0.916 & $<0.001$ & 0.915 & $<0.001$ & 0.753 & $<0.001$ \\
	    Hub & 0.819 & $<0.001$ & 0.900 & $<0.001$ & 0.738 & $<0.001$ \\
	    Authority & 0.911 & $<0.001$ & 0.882 & $<0.001$ & 0.717 & $<0.001$\\
		\bottomrule
	\end{tabular}
	\vspace{-6mm}
\end{table}

Third, \textbf{the labour flow intention implies potential regional brain drain.} All the cities in Heilongjiang and Jilin provinces are \emph{Volcanoes}, while other provinces usually have some \emph{Black holes}. Worse still, their capital cities (i.e., Harbin and Changchun) are among their largest \emph{Volcanoes}, implying significant brain drain in Northeast China~\cite{zhou2018high}. Nevertheless, Liaoning province, another province in Northeast China, has two \emph{Black holes} (i.e., Shenyang and Dalian) to gather labours.  
Meanwhile, we find that cities in big urban agglomerations might also meet brain drain. Specifically, Yangtze River Delta Urban Agglomeration (YRDUA), Pearl River Delta Urban Agglomeration (PRDUA), and Beijing-Tianji-Hebei Urban Agglomeration (BTHUA) are three major urban agglomerations in China~\cite{wei2009comparative} . While \emph{Black holes} usually have scattered spatial distribution in China, these urban agglomerations have \emph{Black hole} groups with 4, 3, and 2  \emph{Black holes}, respectively. In these areas, a few large \emph{Volcanoes} (e.g., Shaoxing in YRDUA, Dongguan in PRDUA, and Langfang in BTHUA) may supply labours for many \emph{Black holes} and easily suffer a brain drain. In addition, Chengdu-Chongqing area forms a \emph{Black hole} group in Southwest China, which is known as the Chengyu urban agglomeration~\cite{chen2020effects}. As this area develops further, nearby cities should pay attention to prevent the potential brain drain, especially Leshan, which is currently the largest \emph{Volcano} within.
\subsection{Assessing the dynamics of labour flow intention during the pandemic}
Here, we use the \emph{Authority} and \emph{Hub} as major dynamic indicators to analyze the change of labour flow intention during the pandemic. The reason is that the absolute \emph{Inflow} and \emph{Outflow} score can be influenced by many factors (i.e., promotion of the search engine) and cannot reflect the real trend of labour flow intention. In contrast, \emph{Authority} and \emph{Hub} scores are based on normalized global link analysis that are robust to reveal the periodicity of labour flow intention. Specifically, for new tier 1 and tier 2 cities, \emph{Authority} scores increased both from 2019Q4 to 2020Q1 and from 2020Q4 to 2021Q1, and decreased both from 2020Q1 to 2020Q2 and from 2021Q1 to 2021Q2, while the \emph{Hub} scores show contrary trend. For lower tiers, the periodicity has a converse shape from higher tiers. Indeed, as labours return to hometowns for spring festival in Q1, flow intentions from small cities to big cities increase, implying a large labour flow after the holiday. Based on the periodicity, we can capture the changes of regional labour attractiveness during different periods of the pandemic, including lockdown (2020Q1), wide reopen (2020Q2), and recovery under regular pandemic control (after 2020Q2). Our analysis shows that the regional labour markets suffered from dramatic changes and demonstrated unusual signs of recovery during the pandemic.

First, \textbf{the labour flow intention deviated from large cities to small cities during the pandemic.} Over the year, the \emph{Authority} scores of tier 1 cities decreased while tier 3 to tier 5 cities increased. Considering the periodicity, we further assess the increase ratio of \emph{Authority} score and \emph{Hub} score from 2019Q4 to 2020Q4. Specifically, the \emph{Authority} scores decreased by 10\% by median in tier 1 cities. In contrast, they increased significantly in cities below tier 2. For new tier 1 and tier 2 cities, although the median change is nearby 0, they generally have positive distribution. 
Different from \emph{Authority}, although there were sharp fluctuations during the peak of the pandemic (i.e., 2020Q1), there shows no obvious annual change of \emph{Hub} scores. In particular, the increase ratio from 2019Q4 to 2020Q4 distribute around 0. This implies that, although dramatically influenced by the pandemic in the short term, labour exportation generally recovered by the end of 2020. Nevertheless, labour flow intention from higher tiers to lower tiers significantly increased.

Second, \textbf{the labour flow intention deviated towards south during the pandemic.} We can observe that labour flow intention generally decreased in North China. Especially, the \emph{Authority} scores of all the major cities in Bohai rim~\cite{dang2020does} decreased, including Beijing (20.5\%), Qingdao (10.9\%), Yantai (7.9\%), Dalian (7\%), Shijiazhuang (6.9\%), Tianjin (6.8\%), Jinan (3.0\%), and Shenyang (2.0\%). Moreover, the \emph{Hub} scores decreased throughout North China. Considering the correlation between labour flow intention and regional development, this may not be a good sign for economic development in Northern China. In contrast, labour flow intention increased in South China. In particular, apart from the small cities, the \emph{Authority} scores of new tier 1 cities also increased, such as Changsha (16.9\%), Chengdu (10.5\%), and Foshan (7.1\%). Moreover, the \emph{Hub} scores increased in South China, which even radiated northwestern areas (e.g., cities in Shaanxi province). These observations imply the general increase of labour flow activeness in South China.

Third, \textbf{the labour flow intention sharply fluctuated during the peak of the pandemic, while high flow intention appeared from small cities to nearby big cities.} During 2020Q1, since the Chinese government took a strict pandemic control policy, the labour market has been dramatically influenced. As labour flow intention in most cities except for tier 1 largely recovered by the end of 2020, we compare the labour flow intention in 2020Q1 with 2021Q1. In small cities (i.e., below tier 2), \emph{Hub} scores increased while \emph{Authority} scores decreased. In contrast, in big cities, \emph{Authority} scores increased while \emph{Hub} scores decreased. Indeed, compared with 2021Q1, flow intention from small cities (i.e., lower than tier 2) to large cities significantly increased in 2020Q1. Especially, new tier 1 suffered the sharpest fluctuation in \emph{Authority} scores (i.e., 10\% higher than 2021Q1). In contrast, the \emph{Authority} scores of tier 1 did not increase much from 2019Q4 to 2020Q1, implying labours paid extra attention to their nearby big cities instead of the super cities.

Fourth, \textbf{the labour flow intention quickly recovered from 2020Q2, while it is faster in South China than North China.} During 2020Q2, the Chinese government gradually lifted the lockdown policy as the pandemic got controlled. Similar to 2020Q1, we also assessed the deviation of labour flow intention in 2020Q2 by comparing the same quarter in the next year (i.e., 2021Q2). Specifically, the median deviation of \emph{Authority} scores quickly dropped to nearby 0 during this period, especially for new tier 1 and tier 2 cities. The difference of \emph{Hub} scores still significantly decreased for new tier 1 (above 10\%) and tier 2 (above 15\%) cities, although it also became closer to 0 than 2020Q1, implying that the labour exportation needed a longer time to recover.
Notably, labour flow intention recovered faster in southern areas than northern areas of China. Specifically, the increase of \emph{Authority} scores was centralized in PRDUA and gradually decayed to further areas, causing a general \emph{Authority} increase in Guangdong province, both including typical big cities such as Shenzhen (40\%), Guangzhou (40\%), Dongguan (46.5\%), and Foshan (60\%), and small cities such as Zhongshan (47.9\%), Huizhou (38.5\%), and Qingyuan (39.0\%). Accompanying with the increase of \emph{Authority}, the \emph{Hub} also burst in the southern areas, implying strong labour exportation strength. In contrast, the \emph{Authority} and \emph{Hub} scores were still low in North China, implying a slower recovery of labour flow intention. In particular, while \emph{Authority} scores increased in some northern cities annually, they still showed a general decrease of \emph{Authority} in 2020Q2.

\subsection{Assessing the dynamics of labour demand during the pandemic}
Here, we analyze the temporal changes of labour demand in different cities, through counting the number of job postings published in different regional labour markets. To avoid the semantic gap in different textual descriptions, we first classified job postings into representative labour demands (see details in Methods), such as blue and white collar. We can observe that regional labour demand structure has largely changed during the pandemic.


First, \textbf{the pandemic sharply decreased the labour demand of blue collar workers relative to that of white collar.} From 2019Q4 to 2020Q4, the job postings for white collars increased by 18.3\% for the whole country, 13.5\% for tier 1, 29.5\% for new tier 1, and 23.0\% for tier 2. Indeed, even during the lockdown period (i.e., 2020Q1), the demand of white collars still increased. Compared to white collars that have stable jobs and can work at home, blue collars got a more severe influence. Specifically, from 2019Q4 to 2020Q4, the demand of blue collars decreased by 60\% for the whole country, 29.1\% for tier 1, 44.4\% for new tier 1, 55.8\% for tier 2, 59.6\% for tier 3, 63.3\% for tier 4, and 67.9\% for tier 5.

Second, \textbf{the demand structure of blue collar jobs changed from manufacturing to service industries during the pandemic.} Specifically, we find that the manufacturing related labour demand got the most severe influence and continuously decreased throughout the year. From 2019Q4 to 2020Q4, the demand decreased by 80.1\% in the whole country, 64.9\% for tier 1, 77.3\% for new tier 1, 82.1\% for tier 2, 80.5\% for tier 3, 80.7\% for tier 4, and 82.3\% for tier 5. Especially, the decrease from 2020Q1 to 2020Q2 was the fastest, with a ratio of 60\% for the whole country. Indeed, many factories broke down during the pandemic in China. Although the decrease slowed down after 2020Q3, the demand remained low and showed no significant increase till 2021Q2. This implies Chinese manufacturing industry needs a longer time to recover after the pandemic. In contrast, express related labour demand had a burst with the reopen policy and changed less from the annual view. As a life-service industry, express quickly recovered after the lockdown period, especially in big cities. This led to the burst of express-related labour demand. Specifically, from 2020Q1 to 2020Q2, the express related labour demand increased by 106\% for the whole country, 172\% for tier 1, 179\% new tier 1, 51.8\% for tier 2, 52.0\% for tier 3, 57.5\% for tier 4, and 46.7\% for tier 5. From 2019Q4 to 2020Q4, the express related labour demand decreased by only 17.4\% for the whole country, 25.2\% for tier 1, 31.4\% for new tier 1, 13.6\% for tier 2, 13.7\% for tier 3, 9.6\% for tier 4, and 4.1\% for tier 5. Different from other blue collar demands, the passenger transport related demands also increased during the pandemic. Especially, from 2019Q4 to 2021Q2, this demand increased by 181.2\% for tier 1, 129.3\% for new tier 1, 139.4\% for tier 2, 171.2\% for tier 3, 151.2\% for tier 4, 98.3\% for tier 5, and 141.0\% for the whole country. This largely raised the whole blue collar demand, especially for small cities. Indeed, the blue collar related labour demand started recovering in 2021 (with 57.1\% in the whole country till 2020Q2), where about 50\% were covered by passenger transport-related jobs.
As a result, proportion of manufacturing continuously decreased in blue collar demands, while ratio of service continuously increased.
Specifically, in 2019Q4, the manufacture, express, and passenger transport jobs occupied 72.7\%, 14.2\%, and 13.1\% of the blue collar demands in the whole country, respectively. In 2021Q2, they became 29.4\%, 20.3\%, and 50.3\%, respectively.


Third, \textbf{the pandemic had different impact on cities with different industrial structures and blue collar demands.} Blue collar demand in big cities recovered fast with the reopening policy. Specifically, from 2020Q1 to 2020Q2, it increased by 151.0\% and 74.7\% for tier 1 and new tier 1 cities, respectively. In contrast, it decreased 17.1\%, 30.9\%, 38.6\%, and 46.3\% for tier 2, tier 3, tier 4, and tier 5 cities, respectively.
The reason is that bigger cities usually have higher proportion of express jobs in their blue collar demands. Indeed, according to our analysis, in 2019Q4, express demand occupied 69.1\%, 36.2\%, and 17.3\% in blue collar demands of the tier 1, new tier 1, and tier 2 cities, while less than 10\% for lower tier cities.
In contrast, small cities had a higher proportion (above 70\% in 2019Q4) of manufacturing related labour demands, which are significantly higher than tier 1 (23.2\%) and new tier 1 cities (49.8\%). Therefore, the pandemic's big impact on manufacturing industry has made blue collar demand difficult to recover in small cities. 
As a result, in the short term, a larger proportion blue collar demand are occupied by big cities. Specifically, in 2020Q2, the tier 2 and above cities occupied 14.6\% of blue collar demands, which is nearly four times of it in 2019Q4 (3.7\%). During this period, big cities quickly increased their attractiveness for blue collars. Nevertheless, with the decrease of express related labour demand in big cities from 2020Q3 and the increase of passenger-transport related labour demand in small cities from 2021Q1, the proportion of different tiers in blue collar demands kept recovering.

Fourth, \textbf{the distribution of labour demand in China showed geographical deviation.} From 2019Q4 to 2020Q4, white collar demands deviated to South China while blue collar demands deviated to North China.
In particular, the white collar demands in the Bohai rim (especially BTHUA) significantly decreased, especially in Beijing (15.2\%). In contrast, the southern urban agglomerations (i.e., PRDUA and YRDUA) increased their white collar demands. In these areas, the demand of many smaller cities increased more than that of super cities. For example, Suzhou (71.7\%), Wuxi (84.3\%), Dongguan (126\%), and Foshan (95\%) increased more than Shanghai (28.1\%), Guangzhou (22.1\%), and Shenzhen (29.8\%). This caused South China to occupy a higher ratio of white collar demands. Although blue collar demands decreased all over China, it is slower in North China than in South China, especially Henan, Shandong, Heilongjiang, and Hebei provinces. Then, North China occupies a higher ratio of blue collar demands.

%% file: data/discussion.tex

\section{Discussion}

To get more insights into the geographical change of labour flow intention, we further discuss the labour demand change in major urban agglomerations during the pandemic.

On white collar labour demands, PRDUA and YRDUA had a similar trend, with two peaks in 2020Q1 and 2020Q3. Indeed, Q1 and Q3 are usually white collar recruitment seasons. This implies that, during the peak of the pandemic, companies in southern urban agglomerations were still working well and had regular recruitment. However, BTHUA only had one peak in 2020Q3. 

This implies that BTHUA missed the spring recruitment in 2020Q1 for the tight lockdown policy. This might cause the white collars to flow to YRDUA and PRDUA. On blue collar demand, YRDUA got the most severe impact. Indeed, among the three agglomerations, YRDUA has the most blue collar demands, which was over twice than those of BTHUA and PRDUA in 2019Q4. Worse still, according to our analysis, nearly 70\% its blue collar demands were manufacturing. Indeed, as the core base of manufacturing industry in China, YRDUA has a long history of developing heavy production entailing metals, automobile, electrical equipment, and machinery~\cite{feng2019relationship}.
With the pandemic's big influence on manufacture, there was a dramatic drop in blue collar demands in YRDUA. As a result, YRDUA largely lost its previous high attractiveness for blue collars. In contrast, blue collar demand in PRDUA depends less on manufacturing (i.e., less than 55\% in 2019Q4), of which a large portion (i.e., above 35\% in 2019Q4) is express. In 2020Q2, the burst of express demand made up the loss of blue collar demand in manufacturing. Manufacturing workers who lost their job can find a job in the express industry. Moreover, potentially related to the less strict pandemic control policy, PRDUA burst much more express demands than the others. Specifically, PRDUA originally has a similar number of express demands with BTHUA. However, from 2019Q4 to 2020Q2, express demand in PRDUA increased by over four times, while BTHUA and YRDUA only doubled. This caused the burst of blue collar demand in PRDUA. It thus attracted a large number of blue collar workers' flow intention. According to the above observations, PRDUA got the least impact on both white collar and blue collar demands, both for political reasons and its industrial structures. As a result, PRDUA gained much attractiveness for labours in the southern and western regions of China. This can be an important reason for the long-term geographical deviation of labour flow intention in China.

In summary, our findings revealed that the pandemic might cause varied impacts on regions with different structures of labour demand and control policies. This analysis can provide timely information for both individuals and organizations in confronting the uncertainty brought by the change of policy and society in the post-pandemic times, and meanwhile benefits governments for better real-time policy-making in the macro regulation of labour market sustainable development.

%% file: data/methods.tex
\section{Methods}
In this part, we introduce the methods used in this paper.

\subsection{Data}
The data used in this paper mainly contain two parts:
    (1) \textbf{Online search query logs.} We use anonymous query logs collected from the most popular search engine in China, across a time span ranging from October 2019 to June 2021. Each record consists of timestamp, location coordinate, query text, and url of the clicked web page. Through the url, we collected the titles of the clicked web pages. 
    (2) \textbf{Job posting data.} The job posting data are collected from major Chinese online recruitment platforms. The time span ranges from October, 2019 to June, 2021. Each record consists of publish timestamp, working city, title, and job description. 
    
\subsubsection{Data Pre-Processing}
We filter job search queries and recognize their origin, destination, and time information as follows: 

\textbf{Job Search Query Filtering.} We filter queries whose contents or clicked titles contain Chinese words ``recruitment'' or ``job hunting''. Then, we deleted duplicated queries.

\textbf{Search Location Mapping.}
To recognize the origin city of the flow intention, we map the geographic coordinate into cities. Specifically, we regard cities as polygons and find out the one that contains the coordinate. We judge if a point is in a polygon with computation geometry algorithms in Shapely~\cite{shapely} package. 


\textbf{Destination Mapping.}
To find out the destination cities, we recognize the places users mentioned or clicked. Specifically, we first build a dictionary containing the official names and alias of Chinese provinces, cities, and districts. 
Then, we match the texts with words in the dictionary using Aho-Corasick algorithm~\cite{aho1975efficient}. This algorithm supports fast multi-pattern string matching with $O(N)$ time complexity, where $N$ denotes the length of the text. 
In particular, some words may mean multiple places. For example, Chaoyang might mean Chaoyang district in Beijing or Chaoyang city in Liaoning province. Therefore, we set three rules to decide the priority of cities for each query: (1) The place that locates in the same province as the origin has higher priority. (2) The place with higher administrative level has 2nd higher priority. (3) The place that is closer to the origin has 3rd higher priority. We check the three rules until we can decide the priority.
After assigning each appeared word with a place, if there are still multiple mentioned places, we choose the one with the lowest administrative level.

\textbf{Cross-City Query Filtering.}
To recognize cross-city labour flow intentions, we drop queries that cannot indicate city-level destination. In particular, district-level places are mapped into their cities. Then, the queries whose origins and destinations are different indicate cross-city labour flow intentions.

Due to large-scaled data, we implemented these operations with Apache Spark~\cite{meng2016mllib} distributed computing framework. After all the above pre-processings, we obtained over 200 millions of cross-city flow intention records in total.

\subsection{Job Posting Clustering}
We recognize the labour demand of the job postings with a textual-based clustering method.

\subsubsection{Keyword Dictionary Generation.}
We use jieba~\cite{sun2012jieba}, a well-known Chinese word cutting tool, to cut job titles into words. After dropping descriptive words and single-character words, we drop low-frequency (less than 1,000) words and high-frequency words (top-50). After manually deleting words that are irrelevant with jobs, we obtain a dictionary with 1,159 words. 

\subsubsection{Keyword-based Clustering.}
For each job posting, we build a vector $x^d \in \mathbb{R}^{N}$ to represent the $d$-$th$ job advertisement, where $N$ means the size of the dictionary, the element $x^d_i = \frac{f^d_i}{\sum_k{f^d_k}}$, $f^d_i$ indicates the frequency of the $i$-$th$ keyword in this title. Then, we use KMeans~\cite{bahmani2012scalable,hamalainen2021improving}, a commonly used clustering algorithm, to cluster the vectors. In particular, we use its Spark MLlib~\cite{meng2016mllib} implementation for parallelized clustering.

\subsection{Measurements of Labour Flow Intention}
We measure labour flow intention on labour flow intention graph with degree centrality and HITS~\cite{kleinberg1998authoritative}.

\subsubsection{Labour Flow Intention Graph Formulation}
In each quarter $t$, we build a labour flow graph $G^t=<V^t, E^t, W^t>$, where each node $v \in V^t$ denotes a city, each directed edge $<o, d> \in E^t$ denotes labour flow intention from origin $o$ to destination $d$. In $W^t$, we weight each edge by the number of flow intention records between this origin-destination (OD) pair. Furthermore, we represent the graph with an adjacency matrix $F^t \in \mathbb{R}^{N \times N}.$ If  $<i,j> \in E^t$, $F^t_{i,j}$ equals the edge weight of OD pair $<i,j>$, otherwise $F^t_{i,j}=0$.

\subsubsection{Degree-based Centrality}
Based on the labour flow intention graph, we calculate \emph{Inflow} and  \emph{Outflow} of a city $i$ in quarter $t$ as $in^t(i) = \sum_kF^t_{i,k}$ and $out^t(i) = \sum_kF^t_{k,i}$, respectively.
To better reveal labour gathering, we mine \emph{Black holes} and \emph{Volcanoes}~\cite{hong2015detecting, li2010detecting} on the labour flow intention graph. In particular, \emph{Black holes} are cities with high net \emph{Inflow} $in^t(i) = in^t(i) - out^t(i)$). In contrast, \emph{Volcanoes} are cities with high net \emph{Outflow} $nout^t(i)=-nin^t(i)$. 

\subsubsection{HITS}
\vspace{-1mm}
Multi-hop labour flow may exist across cities. For example, labours from a small city may first flow to medium cities before further flowing to big cities. Therefore, we use HITS~\cite{kleinberg1998authoritative}, a widely-adopted link-based node importance analysis algorithm, to measure labour flow intention considering multi-hop transitions. Specifically, HITS iteratively measure two traits, namely \emph{Authority} and \emph{Hub}, with the basic idea that (1) nodes with high \emph{Hub} links to nodes with high \emph{Authority}; and (2) nodes with high \emph{Authority} are linked by nodes with high \emph{Hub}. In this work, \emph{Authority} reflects the strength of labour attraction from important cities while \emph{Hub} reflects the strength of labour exporting to important cities.
Formally, we normalize $F^t$ by row and form an adjacency matrix $P \in \mathbb{R}^{N \times N}$, where $P_{i,j} = \frac{F_{i,j}}{\sum^{N}_{k=1}F_{i,k}}.$ Each row of $P$ can be regarded as a city's flow-out distribution to other cities. The \emph{Authority} $A \in \mathbb{R}^{N}$ and \emph{Hub} $H \in \mathbb{R}^{N}$ should satisfy $H = PA$ and $A = P^\mathrm{T}H.$

\subsection{Modularity-based Clustering}
Graphs with high modularity have dense inner-cluster node connections but sparse inter-cluster connections. Specifically, it measures the concentration of inner-cluster edges compared with random distribution of edges regardless of clusters. Formally, given a cluster assignment $\mathcal{C}$ of a graph $F$, modularity is defined as
$Q =\frac{1}{2m} \sum_{i, j}(F_{i,j}–\frac{k_ik_j}{2m})\mathbb{I}\{\mathcal{C}(i) == \mathcal{C}(j)\},$
where $m = \sum_{i,j} F_{i,j}$ denotes the summation of edge weights, $k_i = \sum_{j} (F_{i,j} + F_{j,i})$ denotes the degree of the $i$-$th$ node, $\mathcal{C}(i)$ indicates the cluster of node $i$.
Modularity-based clustering~\cite{newman2006modularity} aims to maximize $Q$, so that nodes in the same cluster are densely connected. In this paper, widely-adopted greedy algorithm, Louvain~\cite{blondel2008fast}, for optimization.
In Louvain, two steps are iteratively performed to enlarge modularity. Specifically, the first step is to find local communities by selectively merging neighboring nodes to enlarge modularity. The second step is to update the network by aggregating each community into one node. In particular, a super-parameter named resolution will control the scale of the clusters.

%% file: main.bbl

\begin{thebibliography}{31}


\ifx \showCODEN    \undefined \def \showCODEN     #1{\unskip}     \fi
\ifx \showDOI      \undefined \def \showDOI       #1{#1}\fi
\ifx \showISBNx    \undefined \def \showISBNx     #1{\unskip}     \fi
\ifx \showISBNxiii \undefined \def \showISBNxiii  #1{\unskip}     \fi
\ifx \showISSN     \undefined \def \showISSN      #1{\unskip}     \fi
\ifx \showLCCN     \undefined \def \showLCCN      #1{\unskip}     \fi
\ifx \shownote     \undefined \def \shownote      #1{#1}          \fi
\ifx \showarticletitle \undefined \def \showarticletitle #1{#1}   \fi
\ifx \showURL      \undefined \def \showURL       {\relax}        \fi
\providecommand\bibfield[2]{#2}
\providecommand\bibinfo[2]{#2}
\providecommand\natexlab[1]{#1}
\providecommand\showeprint[2][]{arXiv:#2}

\bibitem[Aho and Corasick(1975)]%
        {aho1975efficient}
\bibfield{author}{\bibinfo{person}{Alfred~V Aho} {and}
  \bibinfo{person}{Margaret~J Corasick}.} \bibinfo{year}{1975}\natexlab{}.
\newblock \showarticletitle{Efficient string matching: an aid to bibliographic
  search}.
\newblock \bibinfo{journal}{\emph{Commun. ACM}} \bibinfo{volume}{18},
  \bibinfo{number}{6} (\bibinfo{year}{1975}), \bibinfo{pages}{333--340}.
\newblock


\bibitem[Albanesi and Kim(2021)]%
        {albanesi2021effects}
\bibfield{author}{\bibinfo{person}{Stefania Albanesi} {and}
  \bibinfo{person}{Jiyeon Kim}.} \bibinfo{year}{2021}\natexlab{}.
\newblock \showarticletitle{Effects of the COVID-19 Recession on the US Labor
  Market: Occupation, Family, and Gender}.
\newblock \bibinfo{journal}{\emph{Journal of Economic Perspectives}}
  \bibinfo{volume}{35}, \bibinfo{number}{3} (\bibinfo{year}{2021}),
  \bibinfo{pages}{3--24}.
\newblock


\bibitem[Bahmani et~al\mbox{.}(2012)]%
        {bahmani2012scalable}
\bibfield{author}{\bibinfo{person}{Bahman Bahmani}, \bibinfo{person}{Benjamin
  Moseley}, \bibinfo{person}{Andrea Vattani}, \bibinfo{person}{Ravi Kumar},
  {and} \bibinfo{person}{Sergei Vassilvitskii}.}
  \bibinfo{year}{2012}\natexlab{}.
\newblock \showarticletitle{Scalable k-means++}.
\newblock \bibinfo{journal}{\emph{arXiv preprint arXiv:1203.6402}}
  (\bibinfo{year}{2012}).
\newblock


\bibitem[Beil et~al\mbox{.}(2002)]%
        {beil2002frequent}
\bibfield{author}{\bibinfo{person}{Florian Beil}, \bibinfo{person}{Martin
  Ester}, {and} \bibinfo{person}{Xiaowei Xu}.} \bibinfo{year}{2002}\natexlab{}.
\newblock \showarticletitle{Frequent term-based text clustering}. In
  \bibinfo{booktitle}{\emph{Proceedings of the eighth ACM SIGKDD international
  conference on Knowledge discovery and data mining}}.
  \bibinfo{pages}{436--442}.
\newblock


\bibitem[Blondel et~al\mbox{.}(2008)]%
        {blondel2008fast}
\bibfield{author}{\bibinfo{person}{Vincent~D Blondel},
  \bibinfo{person}{Jean-Loup Guillaume}, \bibinfo{person}{Renaud Lambiotte},
  {and} \bibinfo{person}{Etienne Lefebvre}.} \bibinfo{year}{2008}\natexlab{}.
\newblock \showarticletitle{Fast unfolding of communities in large networks}.
\newblock \bibinfo{journal}{\emph{Journal of statistical mechanics: theory and
  experiment}} \bibinfo{volume}{2008}, \bibinfo{number}{10}
  (\bibinfo{year}{2008}), \bibinfo{pages}{P10008}.
\newblock


\bibitem[Bonaventura et~al\mbox{.}(2021)]%
        {bonaventura2021predicting}
\bibfield{author}{\bibinfo{person}{Moreno Bonaventura},
  \bibinfo{person}{Luca~Maria Aiello}, \bibinfo{person}{Daniele Quercia}, {and}
  \bibinfo{person}{Vito Latora}.} \bibinfo{year}{2021}\natexlab{}.
\newblock \showarticletitle{Predicting urban innovation from the US Workforce
  Mobility Network}.
\newblock \bibinfo{journal}{\emph{Humanities and Social Sciences
  Communications}} \bibinfo{volume}{8}, \bibinfo{number}{1}
  (\bibinfo{year}{2021}), \bibinfo{pages}{1--9}.
\newblock


\bibitem[Boustan et~al\mbox{.}(2010)]%
        {boustan2010effect}
\bibfield{author}{\bibinfo{person}{Leah~Platt Boustan},
  \bibinfo{person}{Price~V Fishback}, {and} \bibinfo{person}{Shawn Kantor}.}
  \bibinfo{year}{2010}\natexlab{}.
\newblock \showarticletitle{The effect of internal migration on local labor
  markets: American cities during the Great Depression}.
\newblock \bibinfo{journal}{\emph{Journal of Labor Economics}}
  \bibinfo{volume}{28}, \bibinfo{number}{4} (\bibinfo{year}{2010}),
  \bibinfo{pages}{719--746}.
\newblock


\bibitem[Brodeur et~al\mbox{.}(2021)]%
        {brodeur2021literature}
\bibfield{author}{\bibinfo{person}{Abel Brodeur}, \bibinfo{person}{David Gray},
  \bibinfo{person}{Anik Islam}, {and} \bibinfo{person}{Suraiya Bhuiyan}.}
  \bibinfo{year}{2021}\natexlab{}.
\newblock \showarticletitle{A literature review of the economics of COVID-19}.
\newblock \bibinfo{journal}{\emph{Journal of Economic Surveys}}
  \bibinfo{volume}{35}, \bibinfo{number}{4} (\bibinfo{year}{2021}),
  \bibinfo{pages}{1007--1044}.
\newblock


\bibitem[Chen et~al\mbox{.}(2020)]%
        {chen2020effects}
\bibfield{author}{\bibinfo{person}{Yizhong Chen}, \bibinfo{person}{Hongwei Lu},
  \bibinfo{person}{Jing Li}, {and} \bibinfo{person}{Jun Xia}.}
  \bibinfo{year}{2020}\natexlab{}.
\newblock \showarticletitle{Effects of land use cover change on carbon
  emissions and ecosystem services in Chengyu urban agglomeration, China}.
\newblock \bibinfo{journal}{\emph{Stochastic Environmental Research and Risk
  Assessment}}  \bibinfo{volume}{34} (\bibinfo{year}{2020}),
  \bibinfo{pages}{1197--1215}.
\newblock


\bibitem[Cousins et~al\mbox{.}(2020)]%
        {cousins2020regional}
\bibfield{author}{\bibinfo{person}{Henry~C Cousins}, \bibinfo{person}{Clara~C
  Cousins}, \bibinfo{person}{Alon Harris}, {and} \bibinfo{person}{Louis~R
  Pasquale}.} \bibinfo{year}{2020}\natexlab{}.
\newblock \showarticletitle{Regional infoveillance of COVID-19 case rates:
  analysis of search-engine query patterns}.
\newblock \bibinfo{journal}{\emph{Journal of medical internet research}}
  \bibinfo{volume}{22}, \bibinfo{number}{7} (\bibinfo{year}{2020}),
  \bibinfo{pages}{e19483}.
\newblock


\bibitem[Dang et~al\mbox{.}(2020)]%
        {dang2020does}
\bibfield{author}{\bibinfo{person}{Yunxiao Dang}, \bibinfo{person}{Li Chen},
  \bibinfo{person}{Wenzhong Zhang}, \bibinfo{person}{Dan Zheng}, {and}
  \bibinfo{person}{Dongsheng Zhan}.} \bibinfo{year}{2020}\natexlab{}.
\newblock \showarticletitle{How does growing city size affect residents’
  happiness in urban China? A case study of the Bohai rim area}.
\newblock \bibinfo{journal}{\emph{Habitat International}}  \bibinfo{volume}{97}
  (\bibinfo{year}{2020}), \bibinfo{pages}{102120}.
\newblock


\bibitem[Fang and Dewen(2008)]%
        {fang2008impacts}
\bibfield{author}{\bibinfo{person}{Cai Fang} {and} \bibinfo{person}{Wang
  Dewen}.} \bibinfo{year}{2008}\natexlab{}.
\newblock \showarticletitle{Impacts of internal migration on economic growth
  and urban development in China}. In \bibinfo{booktitle}{\emph{Migration and
  Development Within and Across Borders.}} \bibinfo{pages}{245}.
\newblock


\bibitem[Feng et~al\mbox{.}(2019)]%
        {feng2019relationship}
\bibfield{author}{\bibinfo{person}{Delian Feng}, \bibinfo{person}{Qun Chen},
  \bibinfo{person}{Malin Song}, {and} \bibinfo{person}{Lianbiao Cui}.}
  \bibinfo{year}{2019}\natexlab{}.
\newblock \showarticletitle{Relationship between the degree of
  internationalization and performance in manufacturing enterprises of the
  Yangtze river delta region}.
\newblock \bibinfo{journal}{\emph{Emerging Markets Finance and Trade}}
  \bibinfo{volume}{55}, \bibinfo{number}{7} (\bibinfo{year}{2019}),
  \bibinfo{pages}{1455--1471}.
\newblock


\bibitem[Fortin et~al\mbox{.}(2021)]%
        {fortin2021labor}
\bibfield{author}{\bibinfo{person}{Nicole~M Fortin}, \bibinfo{person}{Thomas
  Lemieux}, {and} \bibinfo{person}{Neil Lloyd}.}
  \bibinfo{year}{2021}\natexlab{}.
\newblock \showarticletitle{Labor market institutions and the distribution of
  wages: The role of spillover effects}.
\newblock \bibinfo{journal}{\emph{Journal of Labor Economics}}
  \bibinfo{volume}{39}, \bibinfo{number}{S2} (\bibinfo{year}{2021}),
  \bibinfo{pages}{S369--S412}.
\newblock


\bibitem[Gillies et~al\mbox{.}(07  )]%
        {shapely}
\bibfield{author}{\bibinfo{person}{Sean Gillies} {et~al\mbox{.}}}
  \bibinfo{year}{2007--}\natexlab{}.
\newblock \bibinfo{title}{Shapely: manipulation and analysis of geometric
  objects}.
\newblock
\newblock
\urldef\tempurl%
\url{https://github.com/Toblerity/Shapely}
\showURL{%
\tempurl}


\bibitem[Ginsberg et~al\mbox{.}(2009)]%
        {ginsberg2009detecting}
\bibfield{author}{\bibinfo{person}{Jeremy Ginsberg}, \bibinfo{person}{Matthew~H
  Mohebbi}, \bibinfo{person}{Rajan~S Patel}, \bibinfo{person}{Lynnette
  Brammer}, \bibinfo{person}{Mark~S Smolinski}, {and} \bibinfo{person}{Larry
  Brilliant}.} \bibinfo{year}{2009}\natexlab{}.
\newblock \showarticletitle{Detecting influenza epidemics using search engine
  query data}.
\newblock \bibinfo{journal}{\emph{Nature}} \bibinfo{volume}{457},
  \bibinfo{number}{7232} (\bibinfo{year}{2009}), \bibinfo{pages}{1012--1014}.
\newblock


\bibitem[Guirao et~al\mbox{.}(2017)]%
        {guirao2017high}
\bibfield{author}{\bibinfo{person}{Bego{\~n}a Guirao}, \bibinfo{person}{Antonio
  Lara-Galera}, {and} \bibinfo{person}{Juan~Luis Campa}.}
  \bibinfo{year}{2017}\natexlab{}.
\newblock \showarticletitle{High Speed Rail commuting impacts on labour
  migration: The case of the concentration of metropolis in the Madrid
  functional area}.
\newblock \bibinfo{journal}{\emph{Land Use Policy}}  \bibinfo{volume}{66}
  (\bibinfo{year}{2017}), \bibinfo{pages}{131--140}.
\newblock


\bibitem[H{\"a}m{\"a}l{\"a}inen et~al\mbox{.}(2021)]%
        {hamalainen2021improving}
\bibfield{author}{\bibinfo{person}{Joonas H{\"a}m{\"a}l{\"a}inen},
  \bibinfo{person}{Tommi K{\"a}rkk{\"a}inen}, {and} \bibinfo{person}{Tuomo
  Rossi}.} \bibinfo{year}{2021}\natexlab{}.
\newblock \showarticletitle{Improving scalable K-means++}.
\newblock \bibinfo{journal}{\emph{Algorithms}} \bibinfo{volume}{14},
  \bibinfo{number}{1} (\bibinfo{year}{2021}), \bibinfo{pages}{6}.
\newblock


\bibitem[Hanushek and Kimko(2000)]%
        {hanushek2000schooling}
\bibfield{author}{\bibinfo{person}{Eric~A Hanushek} {and}
  \bibinfo{person}{Dennis~D Kimko}.} \bibinfo{year}{2000}\natexlab{}.
\newblock \showarticletitle{Schooling, labor-force quality, and the growth of
  nations}.
\newblock \bibinfo{journal}{\emph{American economic review}}
  \bibinfo{volume}{90}, \bibinfo{number}{5} (\bibinfo{year}{2000}),
  \bibinfo{pages}{1184--1208}.
\newblock


\bibitem[Hendricks(2002)]%
        {hendricks2002important}
\bibfield{author}{\bibinfo{person}{Lutz Hendricks}.}
  \bibinfo{year}{2002}\natexlab{}.
\newblock \showarticletitle{How important is human capital for development?
  Evidence from immigrant earnings}.
\newblock \bibinfo{journal}{\emph{American Economic Review}}
  \bibinfo{volume}{92}, \bibinfo{number}{1} (\bibinfo{year}{2002}),
  \bibinfo{pages}{198--219}.
\newblock


\bibitem[Hong et~al\mbox{.}(2015)]%
        {hong2015detecting}
\bibfield{author}{\bibinfo{person}{Liang Hong}, \bibinfo{person}{Yu Zheng},
  \bibinfo{person}{Duncan Yung}, \bibinfo{person}{Jingbo Shang}, {and}
  \bibinfo{person}{Lei Zou}.} \bibinfo{year}{2015}\natexlab{}.
\newblock \showarticletitle{Detecting urban black holes based on human mobility
  data}. In \bibinfo{booktitle}{\emph{Proceedings of the 23rd SIGSPATIAL
  International Conference on Advances in Geographic Information Systems}}.
  \bibinfo{pages}{1--10}.
\newblock


\bibitem[Kleinberg et~al\mbox{.}(1998)]%
        {kleinberg1998authoritative}
\bibfield{author}{\bibinfo{person}{Jon~M Kleinberg} {et~al\mbox{.}}}
  \bibinfo{year}{1998}\natexlab{}.
\newblock \showarticletitle{Authoritative sources in a hyperlinked
  environment.}. In \bibinfo{booktitle}{\emph{SODA}},
  Vol.~\bibinfo{volume}{98}. Citeseer, \bibinfo{pages}{668--677}.
\newblock


\bibitem[Lee(2021)]%
        {lee2021impact}
\bibfield{author}{\bibinfo{person}{Yong-Kwan Lee}.}
  \bibinfo{year}{2021}\natexlab{}.
\newblock \showarticletitle{The Impact of COVID-19 on the Working Conditions of
  Wage Workers-Focusing on Differences by Employment Types}.
\newblock \bibinfo{journal}{\emph{Journal of Labour Economics}}
  \bibinfo{volume}{44}, \bibinfo{number}{2} (\bibinfo{year}{2021}),
  \bibinfo{pages}{71--90}.
\newblock


\bibitem[Li et~al\mbox{.}(2010)]%
        {li2010detecting}
\bibfield{author}{\bibinfo{person}{Zhongmou Li}, \bibinfo{person}{Hui Xiong},
  \bibinfo{person}{Yanchi Liu}, {and} \bibinfo{person}{Aoying Zhou}.}
  \bibinfo{year}{2010}\natexlab{}.
\newblock \showarticletitle{Detecting blackhole and volcano patterns in
  directed networks}. In \bibinfo{booktitle}{\emph{2010 IEEE International
  Conference on Data Mining}}. IEEE, \bibinfo{pages}{294--303}.
\newblock


\bibitem[Meng et~al\mbox{.}(2016)]%
        {meng2016mllib}
\bibfield{author}{\bibinfo{person}{Xiangrui Meng}, \bibinfo{person}{Joseph
  Bradley}, \bibinfo{person}{Burak Yavuz}, \bibinfo{person}{Evan Sparks},
  \bibinfo{person}{Shivaram Venkataraman}, \bibinfo{person}{Davies Liu},
  \bibinfo{person}{Jeremy Freeman}, \bibinfo{person}{DB Tsai},
  \bibinfo{person}{Manish Amde}, \bibinfo{person}{Sean Owen}, {et~al\mbox{.}}}
  \bibinfo{year}{2016}\natexlab{}.
\newblock \showarticletitle{Mllib: Machine learning in apache spark}.
\newblock \bibinfo{journal}{\emph{The Journal of Machine Learning Research}}
  \bibinfo{volume}{17}, \bibinfo{number}{1} (\bibinfo{year}{2016}),
  \bibinfo{pages}{1235--1241}.
\newblock


\bibitem[Newman(2006)]%
        {newman2006modularity}
\bibfield{author}{\bibinfo{person}{Mark~EJ Newman}.}
  \bibinfo{year}{2006}\natexlab{}.
\newblock \showarticletitle{Modularity and community structure in networks}.
\newblock \bibinfo{journal}{\emph{Proceedings of the national academy of
  sciences}} \bibinfo{volume}{103}, \bibinfo{number}{23}
  (\bibinfo{year}{2006}), \bibinfo{pages}{8577--8582}.
\newblock


\bibitem[Pries and Rogerson(2005)]%
        {pries2005hiring}
\bibfield{author}{\bibinfo{person}{Michael Pries} {and}
  \bibinfo{person}{Richard Rogerson}.} \bibinfo{year}{2005}\natexlab{}.
\newblock \showarticletitle{Hiring policies, labor market institutions, and
  labor market flows}.
\newblock \bibinfo{journal}{\emph{Journal of Political Economy}}
  \bibinfo{volume}{113}, \bibinfo{number}{4} (\bibinfo{year}{2005}),
  \bibinfo{pages}{811--839}.
\newblock


\bibitem[Schwandt and Von~Wachter(2019)]%
        {schwandt2019unlucky}
\bibfield{author}{\bibinfo{person}{Hannes Schwandt} {and} \bibinfo{person}{Till
  Von~Wachter}.} \bibinfo{year}{2019}\natexlab{}.
\newblock \showarticletitle{Unlucky cohorts: Estimating the long-term effects
  of entering the labor market in a recession in large cross-sectional data
  sets}.
\newblock \bibinfo{journal}{\emph{Journal of Labor Economics}}
  \bibinfo{volume}{37}, \bibinfo{number}{S1} (\bibinfo{year}{2019}),
  \bibinfo{pages}{S161--S198}.
\newblock


\bibitem[Sun(2012)]%
        {sun2012jieba}
\bibfield{author}{\bibinfo{person}{J Sun}.} \bibinfo{year}{2012}\natexlab{}.
\newblock \showarticletitle{Jieba chinese word segmentation tool}.
\newblock \bibinfo{journal}{\emph{Accessed: Jun}}  \bibinfo{volume}{25}
  (\bibinfo{year}{2012}), \bibinfo{pages}{2018}.
\newblock


\bibitem[Wei(2009)]%
        {wei2009comparative}
\bibfield{author}{\bibinfo{person}{Wang Wei}.} \bibinfo{year}{2009}\natexlab{}.
\newblock \showarticletitle{A comparative study on eco-spatial morphological
  features of the three major urban agglomerations in China}. In
  \bibinfo{booktitle}{\emph{Urban Planning Forum}}, Vol.~\bibinfo{volume}{1}.
  \bibinfo{pages}{46--53}.
\newblock


\bibitem[Zhou et~al\mbox{.}(2018)]%
        {zhou2018high}
\bibfield{author}{\bibinfo{person}{Yang Zhou}, \bibinfo{person}{Yuanzhi Guo},
  {and} \bibinfo{person}{Yansui Liu}.} \bibinfo{year}{2018}\natexlab{}.
\newblock \showarticletitle{High-level talent flow and its influence on
  regional unbalanced development in China}.
\newblock \bibinfo{journal}{\emph{Applied geography}}  \bibinfo{volume}{91}
  (\bibinfo{year}{2018}), \bibinfo{pages}{89--98}.
\newblock


\end{thebibliography}
